# Ensuring Spreadsheet Integrity with Model Master

Jocelyn Paine,

Visiting Fellow, Institute of Learning and Research Technology, University of Bristol

http://www.ifs.org.uk/~popx/

**ABSTRACT**

*We have developed the Model Master (MM) language for describing spreadsheets, and tools for converting MM programs to and from spreadsheets. The MM decompiler translates a spreadsheet into an MM program which gives a concise summary of its calculations, layout, and styling. This is valuable when trying to understand spreadsheets one has not seen before, and when checking for errors. The MM compiler goes the other way, translating an MM program into a spreadsheet. This makes possible a new style of development, in which spreadsheets are generated from textual specifications. This can reduce error rates compared to working directly with the raw spreadsheet, and gives important facilities for code reuse. MM programs also offer advantages over Excel files for the interchange of spreadsheets.*

## 1 INTRODUCTION

Spreadsheets are alarmingly error-prone to write. To quote [Panko 2000], "given data from recent field audits, *most* large spreadsheets probably do contain significant errors". The most recent audits he cites found errors in at least 86% of spreadsheets audited. In a 1997 feature entitled "Fatal Addition" [Ward 1997], *New Scientist* reported that 90% of the spreadsheets audited in a study carried out by Coopers and Lybrand were found to have errors. Given the billions of spreadsheets in use, this leaves the worlds of business and finance horribly vulnerable to programming mistakes. Studies show the chances of any given spreadsheet cell containing an error are somewhere between 0.3% and 3%, so that a spreadsheet of only 100 cells has about a 30% chance of having one error or more.

Creating the spreadsheet is not the only problem. Spreadsheets can be difficult to read, and hence to debug and maintain. Look at the example below - a small professionally-written educational model for calculating income elasticities, shown with formulae visible – and make your best attempt to read it:

Firstly, the use of cell locations – `C9`, `C10` - in formulae makes them hard to understand, because the location's name conveys no information about its purpose. This presumably increases the reader's cognitive load, because of the time needed to match locations with their targets, and the need to hold information in short-term memory while doing so. Excel does provide facilities for naming cells, but many developers don't use these, so trying to understand cell locations is a common readability problem.

Secondly, even in such a small spreadsheet, an annoying amount of horizontal scrolling is needed before one can see all the formulae. This is partly because the spreadsheet contains not only the formulae but also a lot of text and empty cells used for layout. User-interface design experts agree that horizontal scrolling should be avoided wherever possible. In a larger spreadsheet, the reader would have to search a considerable amount of screen area, both to find all the formulae, and to match the cell locations with their targets. It seems reasonable to assume that if the reader cannot see the entire spreadsheet in one go, he or she will have to remember parts of it, further adding to cognitive load. As one developer said: "I end up with vast numbers of notes on little bits of paper". If the spreadsheet has more than one worksheet, flicking between them is an extra inconvenience.

A third difficulty, illustrated by the "Lazy Days" example below [Rajalingham et al 2000], is repeated formulae, due to a calculation being replicated over several time periods, departments, or other entities. This example contains identical calculations of total wage and average wage for each type of personnel:

```
Lazy Days Staff Budget Costs 1995-1996
              Staff    Basic    Overtime  Total    Average
              Numbers  Wages £  Wages £   Wages £  Wages £

Managers      1        17700    0         17700    17700.00
Grade 1       3        45540    1400      46940    15646.67
Grade 2       9        122340   2000      124340   13815.56
Grade 3       12       102350   0         102350    8529.17
Grand Totals  25       287930   3400      291330   11653.20
```

There is much repeated structure here, but the fact that it is repeated is not explicit. The columnar layout and the nature of the application make this likely, but the reader needs to examine and compare all the formulae before being sure.

To overcome such problems, we are developing a tool, named Model Master or MM, that "decompiles" spreadsheets, generating concise specifications of their calculations. The example below shows a specification generated by the decompiler for the elasticity model:

```
attributes <
  new_quantity
  old_quantity
  new_real_income
  old_real_income
  demand_change
  real_income_change
  income_elasticity
  good_type
>
where
  demand_change = new_quantity / old_quantity – 1   and
  real_income_change =
    new_real_income / old_real_income – 1   and
  income_elasticity =
    demand_change / real_income_change   and
  good_type =
    if( income_elasticity > 0,
        "So, this product is a normal good.",
        "So, this product is an inferior good."
      )
```

This lists the variables or "attributes" represented by the cells, and the equations relating them. It gives an alternative view of the spreadsheet, which we believe to be valuable when checking for errors and when trying to understand spreadsheets written by other people.

The significant point about these listings is that they can be regarded as programs. Not only can we decompile existing spreadsheets into this form; we could go in the other direction and compile such programs into spreadsheets. We have in fact devised a complete programming language for describing spreadsheets, which we also call MM, and a compiler that translates MM programs into spreadsheets.

What are the benefits of generating spreadsheets from MM programs, rather than writing directly in Excel? One is that typing errors and other mistakes in an MM program are more likely to be detected than errors in the spreadsheet itself. This is because spreadsheets have little redundancy: if the programmer mistypes a cell location, or through a slip of the cursor accidentally types a formula into the wrong cell, the resulting code still means something to the spreadsheet. On the other hand, when programming MM, one works with names such as `balance` and `start_time,` rather than with cell locations. The programmer must give a list of all such names – to "declare" them – when writing the program. So if a typing error is made, `blance` for example, MM can check the name against this list, realising that since it doesn't occur there, there was an error.

As the example above demonstrates, MM programs consist of almost nothing other than lists of variables together with the equations or formulae relating them. So if the variable names are well chosen – `total_wage` rather than `tot_w`, `hours_per_week` and not `hrpwk` – the programs ought to be comprehensible to non-programmers, so long as they understand the business processes being modelled.

Although in some organisations, spreadsheets are regarded as private to their author and not to be checked by anyone else, others may find this useful in managing development.

Another benefit of MM is code reuse. MM programs can be divided up into separate modules. These can be put into program libraries and included in a program just by mentioning the name of the file where they are stored. Authors can therefore build up a central repository of modules, reducing duplicated work and the risk of cut-and-paste errors. We regard this as extremely important, and emphasise that it is not a trivial claim. This feature of MM stems from mathematical techniques used in designing the OBJ family of algebraic specification languages [Goguen and Tracz 2000], and provides the simplest and most comprehensive module system that we know of.

MM makes possible a style of spreadsheet development similar to that practised with conventional languages such as Fortran and Java. In this, the MM program is primary, and the spreadsheet is merely a means of running it and displaying the results: something to be generated whenever one needs to do a calculation, but otherwise to be ignored. This is extremely different from the way in which most Excel programmers work, and we suspect few would be willing to give up the convenience of interactive model construction and testing possible when you work directly with the spreadsheet. However, because of MM's benefits, particularly with respect to code reuse, we believe it important to find out how it can best be used to complement and increase the integrity of this kind of development. We are therefore continuing work on the compiler and decompiler, and on an integrated user interface which allows them to be invoked directly from Excel.

We also believe that MM programs, whether written from scratch or decompiled out of an existing spreadsheet, are a better medium for interchanging spreadsheets than are Excel XLS files. We are investigating MM for this purpose.

In the rest of this paper, we describe the MM language, compiler, and decompiler from the user's point of view, and discuss the benefits of MM as a medium for spreadsheet interchange. We then give some details of the implementation, and finally summarise our progress and the work yet to be done. We start with an account of the language: without this, it is difficult to understand the compiler and decompiler. The main point of the paper is to present MM in enough detail that we may receive worthwhile criticisms and suggestions for improvement.

## 2 THE MM LANGUAGE AND COMPILER

### 2.1 Attributes, Objects, and Equations

MM views the world in terms of attributes which have values, usually numbers or strings, and which are inter-related by equations. Attributes may be single-valued, or may range over regions of time or space, with a value for each point in the region. It is sometimes convenient to think of them as grouped together to make objects. The simple example below is an MM description of a company:

```
attributes <
  incomings [ 1995:2004 ]
  outgoings [ 1995:2004 ]
  profit    [ 1995:2004 ]
>
where
  profit[ all t ] = incomings[ t ] – outgoings[ t ]
```

(We do realise that as a business application, this program is trivial. This is not a case of the "toy language" syndrome, whose sufferers design languages that won't scale up to realistic applications; we are using MM to build real models. But we do need to simplify so that we don't obscure key concepts.)

To generate a spreadsheet, we pass the program text to the MM compiler, which checks for errors, and if none are found, produces an Excel file. Were we writing the spreadsheet directly, we would almost certainly use successive rows to represent successive years, allocating attributes to adjacent columns. This is what the compiler does, heading each column with the identifier used for its attribute. Appendix 1 shows a generated spreadsheet in the queuing simulation example.

### 2.2 Layout

MM separates the appearance of a spreadsheet from its calculations. All MM programs are written in two parts: the variables and equations, and a separate section stipulating how these are to be arranged and styled. It is this section that deals with matters such as cell formats, background colours, the allocation of variables to cells, and whether a variable is to run down a column or along a row. This means that the results of calculations can be displayed in a variety of ways without changing the program. An interesting example of this is discussed in the section below on Multidimensional Tables.

We based the layout language on HTML, the language in which Web pages are written, largely because this is now very well-developed for expressing layout. And of course, many people have written Web pages and will have some knowledge of it. For those who do not, we shall explain the basics. For the purposes of this paper, the important thing about it is that an HTML document consists of plain text interspersed with tags that specify the document's appearance. For example, `<H1>The MM user manual</H1>` creates a main heading; `<P>…</P>` enclose a paragraph; and the tags `<TABLE>…</TABLE>`, `<TR>…</TR>` and `<TD>…</TD>` are used to display tables, where `<TABLE>…</TABLE>` enclose the entire table, `<TR>…</TR>` enclose a row, and `<TD>…</TD>` enclose cells within the row. Our layout language is similar, augmented with special tags that insert attributes into the tables. Two examples are shown in the section on the decompiler.

### 2.3 Code Reuse; Named Constants

Had we stored the program above in a file called `company`, we could reuse it in another program:

```
include "company"
company2 =
  company
where
  incomings[ 1995 ] = 1000  and
  incomings[ all t > 1995 ] = incomings[ t-1 ] * 1.2
```

Here, we have extended `company` by adding an initial value for the 1995 value of `incomings`, and a growth law for the other values. This example illustrates code reuse. It also shows how one equation can be applied to a number of points within an attribute's domain, as with the formula for profit earlier.

Instead, we could add some new attributes:

```
include "company"
constant average_wage = 500
company3 =
  company plus attributes <
              workforce [1995:2004]
           >
where
  outgoings[ all t ] = workforce[ t ] * average_wage
```

In this program, we have given `company` the new attribute `workforce` representing the number of employees. We calculate `outgoings` from it (again, in a terribly naïve way). This also shows that we can define constants to make code more readable.

In models that simulate some process that spans many regions of time or space, with the same thing happening in each region, constants often enable us to change the number of regions by altering just one number. The queuing simulation in Appendix A has two examples. To achieve the same effect on the raw spreadsheet would require a lot of error-prone copying and editing of cell references.

### 2.4 Parameterisation: Defining Templates

In the above, the company's life span is hard-wired into the code. To change it, we would need to do a global find-and-replace of the years 1995 and 2004. We can avoid this by using parameters to our object definitions. Below, our definition of `company` has become `company_template` and acquired two parameters, `T1` and `T2`. These specify the time over which the attributes range. We then reconstruct `company` from this template. The great thing about this is that we can construct general templates and then use them in a whole variety of applications just by providing suitable parameters.

```
company_template( T1:integer, T2:integer ) = attributes <
  incomings [ T1:T2 ]
  outgoings [ T1:T2 ]
  profit    [ T1:T2 ]
>
where
```

```
    profit[ all t ] = incomings[ t ] – outgoings[ t ]
company = company_template( 1995, 2004 )
```

### 2.5 Function Definitions

The equations in MM programs can call built-in Excel functions, as shown by the queuing simulation, which calls `min` and `rand` amongst others. We considered that it might be useful for users to be able to define their own functions in terms of these built-ins. We implemented this in a prototype, but have omitted it from the current version because of difficulties over its interaction with templates.

### 2.6 Ranges as Arguments; Slices

Many of Excel's built-in functions accept cell-range arguments. For example, `min(C4:D10)` finds the smallest number anywhere within the rectangle whose top-left corner is `C4` and whose bottom right is `D10`. We clearly need an equivalent in MM, which we provide by using the `range` operator. So `min( range incomings )` would find the smallest number anywhere in the `incomings` column. Using the word `range` before the attribute name just adds a bit of redundancy to confirm that the user did indeed intend a range and not just one value.

Sometimes, we need to pass only some of an attribute's values. This is done by "slicing" – using subscripts to select a rectangular subregion. For example, `min( range incomings[1995:2000] )`. The queuing simulation does this. It sets up a series of servers (representing shop assistants or similar), with one column for each. Each row represents one customer; the entry for the customer in row R and the server in column S specifies when S could start processing R. For each customer, the simulation needs to search for the server with the earliest start time. It does this by slicing out the row, and then passing all the server-columns in it to `min`. This explains what is happening with `potential_start_time` in the example.

### 2.7 Ranging over Kinds of Employee: a More Realistic Example

From the examples above, you might gain the impression that MM is useful only for applications where all the attributes range over time. This is not so, as demonstrated by the "Lazy Days" spreadsheet shown in the Introduction. Here is the MM program that would produce it (without the data, which we assume to be typed directly into the spreadsheet):

```
base employee_kind =
  { "Managers", "Grade 1", "Grade 2", "Grade 3",
    "Grand Totals" }
attributes <
  staff_numbers  [ employee_kind ]
  basic_wages    [ employee_kind ]
  overtime_wages [ employee_kind ]
  total_wages    [ employee_kind ]
  average_wage   [ employee_kind ]
>
where
  total_wages[ all e ] =
    basic_wages[ e ] + overtime_wages[ e ]   and
  average_wage[ all e ] =
    total_wages[ e ] / staff_numbers[ e ]
```

What we are doing here is to define a domain called `employee_kind`, and to say that our attributes all have one value for each element of this domain. (We call this domain a "base", a word taken from the mathematics underlying MM.) This is analogous to what we did when making attributes range over the years 1995 to 2004, except that we have replaced the year numbers by symbolic values like "Grade 1". To save space, we have not shown the layout definitions that would produce the above spreadsheet, but essentially, they would consist of `<table>`, `<tr>` and `<td>` tags used so as to position the attributes and the headers `Lazy Days Staff Budget Costs 1995–1996`, `Average`, and so on as they appear in the spreadsheet grid.

### 2.8 Multidimensional Tables

There is no reason for attributes to range over only two dimensions. Had we holographic 3-d computer displays rather than flat monitors, we could imagine a cubical spreadsheet which tabulates the yearly costs due to a company's departments: years running down columns, cost categories (overheads, lighting, transport) running from left to right across the front of the display, and department running from front to back, with a sheet for I.T. at the front, one for Personnel behind it, and a third for Sales behind that. Such attributes are easy to declare:

```
base lifespan = [ 1995:2004 ]
base cost_type = { "Overheads", "Lighting", "Transport" }
base department = { "I.T.", "Personnel", "Sales" }
attribute <
  costs : lifespan * cost_type * department
>
```

Even without a cubical display, we can display such a table in a variety of ways. We might have a number of tabbed panels, with a worksheet on each panel and one panel per year, department running down and cost type running across. We could swap cost type and department. Or we could put all the information on one worksheet, stacking successive years one under the other. And so on.

In such situations, to separate presentation from layout is particularly valuable. Excel enables this to be done to a limited extent by using pivot tables. However, these are an ad-hoc trick. We consider that using a separate layout section is the correct way to go, and we are working on notations to make it easy to express how multidimensional attributes are to be displayed. (Technically, we base these on the fact that the mapping from multidimensional attribute base to two-dimensional worksheet surface is a linear transformation. This makes storage allocation uniform, because the storage allocator can represent all such mappings as matrices.)

There are some spreadsheets, for example Storeys [ProFunda page], that are better than Excel for displaying multidimensional tables. MM would be particularly valuable as a front-end to these.

### 2.9 Units and Dimensional Analysis

There is an Oxford story, recounted in [Morris 1978], that when John Keble was looking after the books of Oriel College, his accounts showed an inexplicable deficiency of between £1800 and £1900. It was eventually discovered that he had added the date to the college liabilities. We have spreadsheets now … and they do nothing to prevent such errors. However, a paper on the programming language Algol68 [Cleaveland 1975] indicates a cure. Cleaveland suggests that it should be possible to declare variables to have units such as meters, seconds or pounds sterling. Knowing the properties of the arithmetic operators, a compiler could then check that expressions were dimensionally correct. If we declared A to be seconds and B to be meters, adding A to B would be invalid (adding length to time makes no sense), but multiplying them would be permitted. Hopefully, this would enable a large number of errors to be detected. The mechanism would easily extend to financial programs, those used by Oriel College included.

We have experimented with this in MM, by augmenting the programming language in two ways. Firstly, the user was allowed to declare names for units, such as cm, lb, £K and $. Secondly, he or she could attach these to attributes, telling the compiler for example that costs has units £K. Appendix 2 shows an example of the compiler checking unit errors.

Although the feature is useful, we have left it out of the current version of MM. One reason is that there are technical problems about the way it interacts with templates. Another is that it complicates function definition, because a function that operates on quantities with units must specify what units each argument can have, and the result's units in terms of these. For example, were we to define square(N) as N*N, we would need to specify that if N has units cm, then the result has units cm*cm, and so on. Yet another problem is that consistent use of units is perhaps too verbose. For example, if we declare a constant length: cm = 1 cm we have told MM that it has units cm. Strictly speaking, if we do not give the quantity assigned to it – 1 – a unit too, we have committed an error, since a dimensionless number is not the same as one with a dimension. There is a tension here between conciseness and correctness which we have not yet resolved.

### 3 THE MM DECOMPILER

### 3.1 Program Transformations

Going from spreadsheets to MM is harder than going in the other direction, because there are many different MM programs that could generate any given spreadsheet. For example, consider a trivial spreadsheet with only two occupied cells:

Value               1

This can always be represented as an MM program with two single-valued attributes. But it could also have originated from a program with one attribute ranging over two points, or from a program with no attributes and two headings, or from a program with one attribute and one heading. We compare two of these possibilities below:

```
attributes < a >              attributes < s a >
where a=1                     where s="Value" and
layout                           a=1
<table>                       layout
 <tr>                         <table>
  <td>Value</td>               <tr>
  <td><attr name="a"/></td>     <td><attr name="s"/></td>
 </tr>                          <td><attr name="a"/></td>
</table>                       </tr>
                              </table>
```

Although the decompiler can make some attempt to decide what the user intended – a column of numbers with text above is probably one multi-valued attribute plus a static heading stating its name – it will in general need to ask the user for more information.

The crucial insight here was that any spreadsheet can be trivially rewritten as an MM program, provided that MM allows any attribute to be arranged anywhere in a spreadsheet, at any position relative to another attribute. (If it didn't, there would be some arrangements not expressible in MM). Once we have this program, we can transform it in various ways, using hints provided by the user or gleaned by examining the spreadsheet.

As an example, consider the Staff Numbers column from the "Lazy Days" spreadsheet shown in the Introduction. The trivial equivalent of this in MM (naming attributes after their cells) is this:

```
attributes < b2 b3 b5 b6 b7 b8 b9 >
where
  b2="Staff" and b3="Numbers" and
  b5=1 and b6=3 and b7=9 and b8=12 and b9=25
```

With a hint from the user, or by noticing that the two name cells contain text and are unrelated to any other cells, we can assume them to be headings, and remove them from the list of attributes, moving them into the layout section:

```
attributes < b5 b6 b7 b8 b9 >
where
  b5=1 and b6=3 and b7=9 and b8=12 and b9=25
```

The remaining attributes are all numbers. With another hint from the user, or by assuming that a column with text above it corresponds to one attribute, we could replace b5 to b9 by one multi-valued attribute:

```
attributes < b5b9[1:5] >
where
  b5b9[1]=1 and b5b9[2]=3 and b5b9[3]=9 and b5b9[4]=12 and
  b5b9[5]=25
```

We call this "rebasing".

With yet another user hint, or by using the heuristic that text above a columnar attribute is most likely its name, we could replace b5b9 by a meaningful name:

```
attributes < StaffNumbers[1:5] >
where
  StaffNumbers[1]=1 and StaffNumbers[2]=3 and
  StaffNumbers[3]=9 and StaffNumbers[4]=12 and
  StaffNumbers[5]=25
```

We can do the same with the other columns. Of course, when an attribute is renamed or rebased, we have to transform any equations that refer to it. This may entail replacing names in the equation, but may also require changing or adding subscripts. For example, any reference to `b9` would have to be replaced by `b5b9[5]` and then by `StaffNumbers[5]`.

### 3.2 Pattern-wizards and rolled equations

In the section on compiler implementation, we talk about "unrolling" equations. This is what the compiler does when it takes an equation like `total_wages[ all e ] = basic_wages[ e ] + overtime_wages[ e ]` that applies to more than one point in an attribute and generates one instance of the equation for each of these points, i.e. one for each cell. To produce concise readable programs, the decompiler has to do the reverse: "rolling".

It would be ideal if the decompiler could do this automatically. However, recognising when to is not easy, and for the moment, the user will need to specify rolling and the range over which to roll as another program transformation. Chris Browne, an expert on spreadsheets and their history, has suggested [Chris Browne's Linux spreadsheet page] that we could start collecting common patterns of repetition and incorporating them into "wizards" that search for instances and offer to apply the most appropriate transformation. He has also suggested that the decompiler be used in the same way as people use algebra manipulation programs: we can provide a large repertoire of transformations, but the user may need to apply a lot of trial and error before finding a "best" MM program for a given spreadsheet.

Isakowitz, Schocken and Lucas [Isakowitz et al 1995] describe a decompiler which has a fair amount of success at rolling equations: we have the impression that the spreadsheets they tried had a very repetitive form generated by drag-and-drop copies. We have been told that these occur very frequently, due to the need to replicate calculations across time periods, so when rolling equations, it probably is worth looking for such patterns.

### 3.3 Error-checking

Once the decompiler has generated its equations, we can inspect them for errors, such as cells that should contain formulae but which have actually been hard-coded (section 5 of [Rajalingham et al 2000] under "Hard coding"). This will, we believe, be the decompiler's most important use as an analysis tool.

Because of the tasks the decompiler has to perform, however, it can also display other information. It has to calculate dependencies between cells, and to discover which cells are used but not initialised, or initialised but not used. Once we have this information, displaying it – either inside Excel by changing cell colours, or in a separate analysis window - is not difficult.

The equations and dependency diagrams are also useful to people who will run but never modify a spreadsheet. In this connection, we are using some routines from the decompiler to extract programs from Excel and implement them for the Web, in a project for the Institute of Learning and Research Technology (ILRT) at the University of Bristol. This involves putting a series of economic models onto the Web as an educational resource – part of the Biz/Ed Virtual Learning Arcade - for students of business and economics [Biz/Ed page].

### 4 MM AS A STANDARD FOR SPREADSHEET INTERCHANGE

We are investigating MM as a medium for the interchange of spreadsheets. At present, XLS files are the most common means of transferring but we believe that using MM programs offers significant benefits. To transfer a spreadsheet developed in MM, one would just send the MM program. To transfer one that was not, one would decompile. As we have already mentioned in the Decompiler section, if one is prepared to help the decompiler by suggesting appropriate program transformations, the resulting program should be easier to understand than its parent spreadsheet.

These are the advantages we believe MM files have for interchange:

- Unlike XLS files, MM programs are text files, which are easy to process with general-purpose tools such as editors. This facilitates debugging when writing programs that process them.

- XLS is a complicated format, and descriptions of it are fairly hard to find, whereas the MM language has a simple syntax which we would be happy to publicise. When writing the MM compiler, we took care to write the parser as a separate module, with a well-defined API and clear documentation describing the data structures produced. We could also make this generally available.

- One may sometimes want to process the spreadsheet formulae in other ways than evaluating them: for example, to produce an annotated cross-reference listing for reference documentation. This is easier to do with MM programs than with XLS files, partly because of the complexity of the latter, and partly because MM programs make more of the structure apparent, for example the replication of calculations over different time periods.

- MM completely separates the program from its appearance. The latter can be changed simply by modifying the layout section.

## 5 IMPLEMENTATION

### 5.1 The Compiler

The compiler uses standard methods from compiler technology, but differs from other compilers in that it allocates storage not for a one-dimensional machine memory, but for the two-dimensional, and visible, medium of a spreadsheet. It has 7 passes:

1. Input: reads and parses the input and builds the syntax tree. We use the freeware JavaCC [JavaCC page] parser-generator, which translates a grammar into a Java parser.

2. Semantic checking: analyses the syntax tree for errors such as undeclared identifiers and objects defined in terms of themselves. One theme that occurs here and in the decompiler and Web-based spreadsheet evaluator is the need to discover dependencies between attributes. There are standard directed-acyclic-graph algorithms for this.

3. Identifier substitution: replaces all identifiers by their definitions.

4. Colimit: performs an algebraic operation analogous to multiplying together all the object definitions in the tree, resulting in one big object definition for the entire program.

5. Storage allocation: allocates locations for all attributes, taking them from the layout section if there is one. This generates a map associating each attribute with a cell address.

6. Code generation: "unrolls" equations, for example replacing `profit[ all t ] = incomings[ t ] - outgoings[ t ]` by one equation for each value of t. Then it replaces attributes by their cell addresses. This gives us a "cellmap": a map associating each used cell with the formula it contains.

7. Output: writes out the formulae together with instructions for placing them into their cells.

The compiler is written in Java, for portability. Why worry about this, when Excel only runs under Windows? Firstly, for our Web work, we want to generate code for our own Web-based spreadsheet engine, which may not be running on a Windows server. That will not concern many users, but another point is that Excel is not the only spreadsheet, and we do not want to restrict the range of machines on which MM can be used. Unix, for example, has the public-domain Oleo, Gnumeric and Dismal spreadsheets amongst others (a list can be found at [Spreadsheet FAQ page]), and MM would be as useful with these as with Excel. We have also mentioned its potential for use with ProFunda.

Because of portability, we are committed to Java, but the language has undoubted disadvantages for writing compilers. It is verbose; it is object-oriented to an extent which just does not suit compiler-writing; and there are certain idioms very useful here which Java has trouble with, such as higher order functions. For these reasons, we have written large parts of MM in Kawa [Kawa page], a Java implementation of the functional programming language Scheme. Kawa programs can be linked with Java, and so do not compromise portability.

**Mathematical Background**

MM was developed from a branch of mathematics known as category theory. Here is not the place to discuss details. Briefly stated however, category theory, like logic, is a tool for studying mathematical and computational concepts, one concerned much more with their form than with their content. It comes equipped with very general equivalents of common notions such as sum and product which can be applied to a vast and varied range of different situations. The computer scientist Joseph Goguen [Goguen 1975, Goguen 1992] has put it to use, together with another branch of mathematics known as sheaf semantics, in answering the questions "what is an object?", and "what does it mean to say a system is composed out of objects?". It turns out that his formulation can be implemented as a new style of programming, which we call System Limit Programming –

"limit" being the mathematical operation used to assemble components into a system. This gave us the attribute-and-base notation for MM, and – together with the ideas described in [Goguen and Tracz 2000] – the module structure.

### 5.2 The Decompiler

The decompiler has four passes:

1. Input: reads formulae from a spreadsheet file, again parsing it with JavaCC. Converts these into the same internal representation as that generated by pass 6 of the compiler, and then builds a cell-map.

2. Dependency analysis: builds a graph of cell dependencies, enabling input-only and output-only cells to be identified. An optional phase here checks for text cells with no dependants or dependees and assumes them to be headers and other static text, removing them from the cellmap.

3. Program transformation: reads and obeys transformation commands, rewriting attribute lists and equations as directed.

4. Output: pretty-prints the result as an MM program.

It is interesting to compare the decompiler with the factoring algorithm described by Isakowitz, Schocken and Lucas [Isakowitz et al 1995]. We developed our decompiler independently, but there are similarities. Like us, the authors recommend separating layout, which they call "editorial" information, from program, and their algorithm does so. It works in two stages, first allowing the user to split the spreadsheet into regions corresponding to separate attribute groups (we would call these objects: Isakowitz et al refer to them as relations). We do not do this yet, regarding the spreadsheet as "flat", with all attributes in the same object, but it probably is useful. We are considering identifying blocks automatically, by using a depth-first region-numbering algorithm to find all disconnected regions of the spreadsheet.

Their second stage removes the static text, and then converts each relation into an MM-like language which the authors call FRL. Unlike us, the authors see this just as a convenient internal representation, and not as something to be understood by the user. FRL does not have explicit bases: in our terms, all attributes are based from 1 upwards.

### 6 PROGRESS AND FURTHER WORK

We have built several prototypes of the MM compiler, testing language features such as dimensional analysis, and are now completing a version which uses the same internal representations as, and will be integrated with, the decompiler. For the latter, some work still has to be done on allowing the user to describe program transformations conveniently, and on heuristics for identifying attribute names and automatically acquiring other useful information from the spreadsheet.

At present, both the compiler and decompiler are command-line-driven programs which communicate with Excel via SYLK files [SYLK page]. This is convenient for development, because SYLK is a textual representation, much easier to read and to edit than XLS. However, Microsoft have not bothered to maintain it – one source states that it has not been revised since Excel 2.0 – and it can not represent all the style information used in later versions of Excel. Nor can it handle multiple worksheets. To overcome this, we intend to connect MM directly to Excel so that it can be run as an Excel add-on, taking information from, and passing it to, Excel cells and cell ranges. This will also make it possible to provide the user with an integrated development environment for MM, which we regard as very important.

The largest model on which we have tested the compiler is the queuing simulation below; the decompiler has so far been run only on very small spreadsheets, of similar size to the examples in the introduction. We shall be trying it out on more spreadsheets of this kind when putting them onto the Web, as we describe under the decompiler section on Error Checking. This will be the only usability evaluation under the current funding, though we hope to set up a joint project with Clemson University to do more formal evaluation.

For the longer term, we note as we did in the introduction that MM enables two kinds of spreadsheet development. One is the conventional kind where one works directly with the spreadsheet. Here, MM is useful just as an analysis and error-detection tool, through its decompiler. The other is similar to that practised with conventional languages such as Fortran and Java. In this, the MM program is primary, and the spreadsheet is merely a means of running it and displaying the results: something to be generated whenever one needs to do a calculation, but otherwise to be ignored. Using MM in this way, one can take full advantage of the facilities for code reuse, and can easily restructure spreadsheets as described under the section on Multidimensional Tables. We need to find out how best to combine the two kinds of development so as to provide the convenience of interactive inputting and testing with the readability, integrity and maintainability of MM programs.

## 7 ACKNOWLEDGEMENTS

I have done most of MM's design and implementation in my own time. However, funding for putting the models onto the Web, and for the latest work on the compiler and decompiler was provided via the ILRT by the Joint Information Systems Committee 5/99 call for teaching and learning resources. I wish to thank Jack Ponton of the Department of Chemical Engineering at the University of Edinburgh for telling me about SYLK files, Margaret van Biene-Hershey for a valuable discussion on spreadsheet integrity as well as a delightful visit to Harmelen, Rob Kemmeren, and the anonymous referees for their comments on the draft version of this paper.

## APPENDIX 1: QUEUING SIMULATION IN MM

The program below is an MM version of a queuing simulation. We designed it by reverse-engineering a spreadsheet written by Thomas Grossman, a professor of management science at Calgary University. He has written a variety of simulations in order to help students understand queuing behaviour [Grossman 1999]. Ours is derived from his basic four-server queue, downloadable as `4vanilla.xls` from the Web page cited with this paper. Because the original spreadsheet, while interesting, is relatively complex and difficult to understand, we thought it was a good test for MM.

The program demonstrates several features of MM.

- Comments for program documentation.

- Attributes ranging over more than one dimension. The attribute `potential_start_time` has one dimension representing time: this runs from row to row, as with the other attributes. Its other dimension represents servers, and runs along the columns: there is one column for each server.

- Constant declarations make changing the spreadsheet structure easy. `potential_start_time` is declared as ranging over `event*[1:N]`, where `event` is itself declared as `[1:10]`. In other words, `potential_start_time` ranges over the two-dimensional region `[1:10,1:4]`. The N referred to here is declared as a constant equal to 4. Simply changing this to another value would automatically change the number of columns allocated to the servers. Consider how little effort this is compared to modifying the spreadsheet itself!

- The use of Excel's built-in functions `if`, `match`, `min` and `rand`.

- Cell ranges in function arguments. In the function call `min( range potential_start_time[e] )`, `e` is time: it selects one of the rows. `potential_start_time` has one column for each server, so the expression `range potential_start_time[e]` causes a cell range containing all these columns within the row to be passed to `min`.

- Commands for setting cell formats, and for placing headings in columns. We have now replaced these by the HTML-based layout language. Here is the program:

```
/* Queue.mm */

This is a process-driven simulation of queuing. It has two main components: a
set of customers, and a set of servers. The number of customers and servers is
fixed before starting. Customers enter a shop, go to a server, queue, get
served, and leave. Each customer interacts once only and then leaves.

We could slice up time by allocating equally-spaced time points to successive
rows. Instead, following Grossman, each row represents one customer's complete
sequence of transactions: arriving, starting to queue, being served, and
leaving. We handle this in MM by declaring the relevant attributes to run over a
base (domain of observations) 1:10 similar, where the e'th point represents the
e'th customer. These attributes are, for example, customer_number, arrival_time,
service_start_time and service_time, all quantified over 'event'.

Again following Grossman, we allocate one column to represent each server. The
only attribute of a server that we use directly is the potential start time that
the customers would be served by it. So the columns hold these times. In MM, we
do this by declaring the attribute potential_start_time's base to have a second
dimension, running from 1 to N, the number of servers.
```

```
By default, the MM compiler puts the name of each attribute above its column.
You can change this with the 'name' qualifier. The 'format' qualifier changes
the format of the cells for the attribute values.
*/
   base event = [1:10];
   // Each point represents one customer.
   constant N = 4;
   // The number of servers.
   < customer_number : event
                       name "Customer" br "#"
     // The e'th element is the number of the e'th customer.
     interarrival_time : event
                         name "Interarrival" br "duration"
     // The e'th element is the time between the arrivals of
     // the e-1'th and e'th customers. For customer 1, it's
     // the time between start and the customer's arrival.
     interarrival_time_mins : event
                              name "Interarrival" br
                                   "duration" br "(mins)"
                              format 0.00
     // The same thing but in minutes, for display.
     arrival_time : event
                    name "Arrival"
                    format hh:mm
     // The e'th element is the time at which the e'th
     // customer arrives.
     potential_start_time : event * [1:N]
                            name "Potential" br "start"
                            format hh:mm
     // The e,N'th element is the time at which the e'th
     // customer could start being served by server N, given
     // that it may already be busy serving someone else.
     next_server : event
                   name "Server" br "#"
     // The e'th element is the server that will be used for
     // the e'th customer.
     service_start_time : event
                          name "Service" br "start"
                          format hh:mm
     // The e'th element is the time at which the e'th
     // customer starts being served.
     service_end_time : event
                        name "Service" br "end"
                        format hh:mm
     // The e'th element is the time at which the e'th
     // customer finishes being served.
     service_time : event
                    name "Service" br "duration"
     // The e'th element is the time taken to serve the e'th
     // customer.
     service_time_mins : event
                         name "Service" br "duration" br
                              "(mins)"
                         format 0.00
     // The same thing but in minutes, for display.
     start name "Start"
           format hh:mm
     // When the simulation begins (time at which the servers
   // open).
>

where

    customer_number[all e] = e   and
    // Every customer has a unique number. We display this,
```

```
       // but don't use it in the calculation.

    interarrival_time[all e] = 10 * rand() / ( 24 * 60 )   and
    // The interarrival times are taken from a uniform
    // distribution over the interval 0..10 minutes. Excel
    // represents times as 1 day = 1, and its rand() function
    // returns numbers from the interval 0.0..1.0, so we
    // scale down by dividing by 24*60 to bring the range to
    // 1 minute, then multiplying by 10.
    interarrival_time_mins[all e] =
      interarrival_time[e] * 24*60   and
    // Scale to 1 minute = 1, for nice display.
    service_time[all e] = 20 * rand() / ( 24 * 60 )   and
    // Service times are generated in the same way.
    service_time_mins[all e] = service_time[e] * 24*60   and
    // Scale to 1 minute = 1, for nice display.
    arrival_time[all e>1] =
      arrival_time[e-1] + interarrival_time[e]   and
    arrival_time[1] = start + interarrival_time[1]   and
    // The arrival time of each customer is the arrival time
    // of the previous one (or the start time, for customer
    // 1) plus the generated interarrival time.
    potential_start_time[all e>1,all N] =
      if( next_server[e-1] = N,
          service_end_time[e-1], arrival_time[e] )   and
    potential_start_time[1,all N] = start   and
    // potential_start_time[e,N] is the earliest time at
    // which server N can serve customer e, given that it may
    // already be busy. For all servers,
    // the potential start time for the first customer is the
    // start of the simulation. The potential start time for
    // other customers is the time at which the
    // server finishes with the previous customer, if it has
    // one; otherwise the customer's arrival time.
    service_start_time[all e] =
      min( range potential_start_time[e] )   and
    // The time when we can actually start is the minimum of
    // the potential start times. We calculate this by using
    // Excel's min function. The MM 'range' construct
    // delivers the range of potential_start_time
    // across all servers.
    service_end_time[all e] =
      service_start_time[e] + service_time[e]   and
    // The time when the customer finishes being served is
    // their start time plus the generated service time.
    next_server[all e] =
      match( service_start_time[e],
             range potential_start_time[e], 0 )   and
    // To find the first free server for customer e, we use
    // Excel's match function. This scans the potential start
    // times until it finds service_start_time, and returns
    // the index.
    start = 9 / 24
  // The start time is 9:00 am.
```

This is the spreadsheet MM generates:

| Customer # | Interarriva duration | Interarriva duration (mins) | Arrival | Potential start | | | | Server # | Service start | Service end | Service duration | Service duration (mins) | Start |
|---|---|---|---|---|---|---|---|---|---|---|---|---|---|
| 1 | 0.000845 | 1.22 | 09:01 | 09:00 | 09:00 | 09:00 | 09:00 | 1 | 09:00 | 09:02 | 0.001776 | 2.56 | 09:00 |
| 2 | 0.005899 | 8.50 | 09:09 | 09:02 | 09:09 | 09:09 | 09:09 | 1 | 09:02 | 09:10 | 0.005462 | 7.87 | |
| 3 | 0.005184 | 7.46 | 09:17 | 09:10 | 09:17 | 09:17 | 09:17 | 1 | 09:10 | 09:21 | 0.007887 | 11.36 | |
| 4 | 0.001579 | 2.27 | 09:19 | 09:21 | 09:19 | 09:19 | 09:19 | 2 | 09:19 | 09:27 | 0.005732 | 8.25 | |
| 5 | 0.004946 | 7.12 | 09:26 | 09:26 | 09:27 | 09:26 | 09:26 | 1 | 09:26 | 09:29 | 0.002219 | 3.20 | |
| 6 | 0.001618 | 2.33 | 09:28 | 09:29 | 09:28 | 09:28 | 09:28 | 2 | 09:28 | 09:34 | 0.004124 | 5.94 | |
| 7 | 0.002856 | 4.11 | 09:33 | 09:33 | 09:34 | 09:33 | 09:33 | 1 | 09:33 | 09:33 | 0.000268 | 0.39 | |
| 8 | 0.001633 | 2.35 | 09:35 | 09:33 | 09:35 | 09:35 | 09:35 | 1 | 09:33 | 09:52 | 0.013551 | 19.51 | |
| 9 | 0.001845 | 2.66 | 09:38 | 09:52 | 09:38 | 09:38 | 09:38 | 2 | 09:38 | 09:45 | 0.004895 | 7.05 | |
| 10 | 0.005995 | 8.63 | 09:46 | 09:46 | 09:45 | 09:46 | 09:46 | 2 | 09:45 | 09:57 | 0.008421 | 12.13 | |

## APPENDIX 2: ERROR DETECTION

### A2.1 USE OF IDENTIFIERS

The extract below demonstrates error-checking. We edited the queuing simulation so that it contained a duplicated identifier and two undeclared ones. The compiler has detected these and reported them to the listing file.

```
  1: /* Queue.mm */
  2:
...
 97:
 98:    service_time : event
 99:               name "Service" br "duration"
100:    // The e'th element is the time taken to serve the e'th
101:    // customer.
102:
103:    service_time : event
       ^
    Error: Duplicate attribute service_time in object
           ...
132:
133:    arrival_time[all e>1] =
134:       arrival_time[e-1] + interarrival_tim[e]   and
                               ^
    Error: Undeclared identifier interarrival_tim
...
139:
140:    potential_start_time[all e>1,all N] =
141:       if( next_server[e-1] = M,
                                  ^
    Error: Undeclared identifier M
```

### A2.2 UNITS

The extract below shows the compiler detecting unit mismatch errors in constant definitions. Two of these are errors across the + operator: the compiler assumes that adding quantities with different units is not allowed. The other is an error in the ^ (power) operator, where it is assumed that one can only raise a quantity to a dimensionless power. So raising a length to the power 2 to get an area is permitted, but raising it to the power of 2 times another length does not make sense.

```
  1: /* UnitsIncompatibleErrors.mm */
  2:
  3: unit cm
  4: unit sec
  5: unit £
  6:
  7: constant a = 1 cm + 2 sec
                     ^
    Error: The left-hand argument has units cm, but the
           right-hand argument has units sec.
...
 15: constant e = (1 cm) ^ (2 cm)
                          ^
    Error: Operator ^ expects something with no units here,
           not units cm.
...
 22:
 23: constant i = 2(cm/sec) + 3cm*1£
                            ^
    Error: The left-hand argument has units cm * sec^-1, but
           the right-hand argument has units cm * £.
```

---